\documentstyle[12pt]{article}

\def\pur#1(#2){\ifx#2@@\nu_{#1}\else\nu_{#1}(#2)\fi}
\def\puri(#1){\pur\infty(#1)}
\def\purS(#1){\pur H(#1)}

\begin{document}

\title{On Some Additivity Problems in Quantum Information Theory}
\author{G. G. Amosov \thanks{
Electronic Mail:{\tt gramos@mail.sitek.ru}} ,\ A. S. Holevo\thanks{
Electronic Mail: {\tt holevo@mi.ras.ru}},\ and R.~F. Werner\thanks{
Electronic Mail: {\tt R.Werner@tu-bs.de}} \\
{\small Moscow Institute for Physics and Technology}\\
{\small Steklov Mathematical Institute, Moscow}\\
{\small Institut f{\"u}r Mathematische Physik, TU Braunschweig}}
\date{\today }
\maketitle

\section{Introduction}

Quantum information theory \cite{ben} is not merely a theoretical basis for
physics of information and computation. It is also a source of challenging
mathematical problems, often having elementary formulation but still
resisting solution. It appears that surprisingly little is known about what
may be called the combinatorial geometry of tensor products of Hilbert
spaces, even in finite dimensions. One group of open problems concerns the
additivity properties of various quantities characterizing quantum channels,
notably the capacity for classical information, and the ``maximal output
purity'', defined below. All known results, including extensive numerical
work in the IBM group \cite{smo}, the Quantum Information group in the
Technical University of Braunschweig, and elsewhere, are consistent with the
conjecture that these quantities are indeed additive (resp.\ multiplicative)
with respect to tensor products of channels. A proof of this conjecture
would have important consequences in quantum information theory: in
particular, according to this conjecture, the classical capacity or the
maximal purity of outputs cannot be increased by using entangled inputs of
the channel.

In this paper we state the additivity/multiplicativity problems, give some
relations between them, and prove some new partial results, which also
support the conjecture.

\section{Statement of the problem}

Let us give precise formulation of the additivity problem for the classical
capacity (see \cite{ben}, \cite{hol}). Let ${\cal B(H)}$ be the $\ast $
-algebra of all operators in a finite dimensional unitary space ${\cal H}$.
We denote the set of states, i.e. positive unit trace operators in ${\cal 
B(H)}$ by ${\cal S(H)}$, the set of all $m$-dimensional projections by $
{\cal P}_{m}({\cal H)}$ and the set of all projections by ${\cal P(H)}$. A
quantum channel $\Phi $ is a completely positive trace preserving linear map
of ${\cal B(H)}$ (we are in the finite dimensional case and we use the
Schr\"odinger picture). These are the maps admitting the Kraus decomposition
(see e. g. \cite{lin}, \cite{hol}) 
\begin{equation}
\Phi (\rho )=\sum_{k}A_{k}\rho A_{k}^{\ast },  \label{kra}
\end{equation}
where $A_k$ are operators satisfying $\sum_{k}A_{k}^{\ast } A_{k}=I.$

Let $H(\rho )=- \mbox{Tr}\rho \log \rho $ denote the von Neumann entropy of
the state $\rho $ and define 
\[
\label{CPhi} C(\Phi )=\max_{p_{i},\rho_{i}}[H(\sum_{i}p_{i}\Phi (\rho
_{i}))-\sum_{i}p_{i}H(\Phi (\rho_{i}))], 
\]
where the maximum is taken over all finite probability distributions $
\{p_{i}\}$ on ${\cal S(H)}$, ascribing probabilities $p_{i}$ to (arbitrary)
states $\rho_{i}$. The quantity $C(\Phi)$ appears as the ``one-step
classical capacity'' of the quantum channel $\Phi$ or the capacity with
unentangled input states (we refer to \cite{hol} for a detailed
information-theoretic discussion and the proof of the corresponding coding
theorem). A thorough discussion of the properties of $C(\Phi)$ is given in 
\cite{sch}.

The additivity problem can be formulated as follows: let $\Phi_{1},\dots ,\Phi
_{n}$ be channels in the algebras ${\cal B(H}_{1}{\cal )},\dots ,{\cal B(H}
_{n}{\cal )}$ and let $\Phi_{1}\otimes \dots \otimes \Phi_{n}$ be their
tensor product in ${\cal B(H}_{1}\otimes \dots \otimes {\cal H}_{n}{\cal )}$
. Is it true that 
\begin{equation}
C(\Phi_{1}\otimes \dots \otimes \Phi_{n})=\sum_{i=1}^{n}C(\Phi_{i})\quad?
\label{add}
\end{equation}
This obviously holds for reversible unitary channels; in \cite{hol} the
additivity was established for the so called classical-quantum and
quantum-classical channels, which map from or into an Abelian subalgebra of
${\cal B(H)}$. 

Another closely related problem is the additivity of a quantity, which can
be read as the ``maximal output purity'' of a channel. In fact, there are
several quantities of this kind, depending on the way we measure ``purity''.
If we just take the von Neumann entropy as a measure of purity, we
arrive at the question \cite{fuchs}, \cite{rusk} whether or not 
\begin{equation}
\min_{\rho \in {\cal S}({\cal H}_{1}\otimes \dots\otimes {\cal H}_{n}) }
H((\Phi_{1}\otimes \dots \otimes \Phi_{n})(\rho )) =\sum_{i=1}^{n}\min_{\rho
\in {\cal S(H}_{i})}H(\Phi_{i}(\rho )) \; ?  \label{ent}
\end{equation}
For a particular class of channels this property implies (\ref{add}) (see
the Lemma in Section~\ref{sec:dep} below).

We will also consider this problem for other measures of purity, based on
the noncommutative $\ell_p$-norms 
\[
\Vert A\Vert_{p}=\left( {\rm Tr}|A|^{p}\right) ^{\frac{1}{p}}\;, 
\]
defined for $p\geq 1$, and $A\in {\cal B(H)}$, with the operator norm $\Vert
A\Vert $ corresponding naturally to the case $p=\infty$. For an arbitrary
quantum channel $\Phi$ let us introduce the following notations for the
``highest purity'' of outputs of a channel 
\begin{eqnarray}
\purS(\Phi ) &=&\min_{\rho }H(\Phi (\rho )), \\
\pur p(\Phi ) &=&\max_{\rho }\Vert \Phi (\rho )\Vert_{p}, \\
\pur{-\infty }(\Phi ) &=&\min_{\rho }\Vert \Phi (\rho )^{-1}\Vert ^{-1},
\label{mi}
\end{eqnarray}
where the extrema are taken with respect to all input density matrices $\rho$
. By convexity of the norms, the extrema in the above definitions are
attained on pure states (in the first (resp. last) case the operator
convexity of the function $x\mapsto x\log x$ (resp. $x\mapsto x^{-1}$) is
also relevant).

Then the additivity/multiplicativity inequalities 
\begin{eqnarray}
\pur p(\Phi_{1}\otimes \Phi_{2}) &\geq &\pur p(\Phi_{1})\pur p(\Phi_{2})
\label{mp} \\
\purS(\Phi_{1}\otimes \Phi_{2}) &\leq &\purS(\Phi_{1})+\purS(\Phi_{1})\;
\label{ae}
\end{eqnarray}
are clear from inserting product density operators into the defining
variational expressions. The standing {\bf conjecture} is that equality
always holds in these inequalities, i.e. that choosing entangled input
states is never helpful for getting purer output states.

Before proceeding to show some new partial results on this problem, it is
helpful to establish the relation between $\purS(\Phi)$, and $\pur p(\Phi)$
for $p$ close to one. Of course, some relationship is expected, as the von
Neumann entropy $H(\rho)$ can be computed in terms of the derivative of $
\Vert\rho\Vert_p$ at $p=1^+$. Here we find that if the equality holds in (
\ref{mp}) for $p$ arbitrarily close to $1$, then it holds also in (\ref{ae}).

{\it Proof. }We shall use the fact that for every $0<x\leq 1$ 
\[
\frac{1-x^{p}}{p-1}\uparrow -x\frac{\log x}{\log e} 
\]
if $p\downarrow 1$. Thus $\frac{1-{\rm Tr}\Phi (\rho )^{p}}{p-1}$ is a
monotonely increasing family of continuous functions of the variable $\rho $
which varies in the compact set ${\cal S(H)}$, converging pointwise to the
continuous function $H(\Phi (\rho ))$. By Dini's Theorem, the convergence is
uniform, and 
\[
\min_{\rho }H(\Phi (\rho ))=\lim_{p\downarrow 1}\frac{1-\max_{\rho }{\rm Tr}
\Phi (\rho )^{p}}{p-1}. 
\]
Therefore, if the equality holds in (\ref{mp}) for $p$ close to $1$, 
\[
\min_{\rho }H(\left( \Phi_{1}\otimes \Phi_{2}\right) (\rho
))=\lim_{p\downarrow 1}\frac{1-\max_{\rho }{\rm Tr}\left( (\Phi_{1}\otimes
\Phi_{2})(\rho )\right) ^{p}}{p-1} 
\]
\[
=\lim_{p\downarrow 1}\frac{1-\max_{\rho }{\rm Tr}\left( \Phi_{1}(\rho
)\right) ^{p}\max_{\rho }{\rm Tr}\left( \Phi_{2}(\rho )\right) ^{p}}{p-1}
=\min_{\rho }H(\Phi_{1}(\rho ))+\min_{\rho }H(\Phi_{2}(\rho )). 
\]
$\Box $

\section{Tensoring with an ideal channel}

The first natural step is to establish the multiplicativity property when
one factor is the identity channel.

{\bf Lemma. }For $\ast =p,H, -\infty$ 
\begin{equation}
\pur*(\Phi \otimes {\rm Id})=\pur*(\Phi )\;.
\end{equation}
Since $\pur p({\rm Id})=1$, and $\purS({\rm Id})=0$, this is indeed an
instance of the additivity/multiplicativity conjecture.

{\it Proof. }We shall restrict to the case{\it \ }$\ast =p, 1\le p\le\infty$
. The argument in the case $\ast =-\infty $ is similar and the case{\it \ }$
\ast =H$ follows by the argument given above.

Let us denote by ${\cal H}_{1},{\cal H}_{2}$ the Hilbert spaces of the first
and the second system, respectively. Let $\phi_{12}$ be a unit vector in $
{\cal H}_{1}\otimes {\cal H}_{2},$ and write $\rho_{12}=|\phi_{12}\rangle
\langle \phi_{12}|$ and $\rho_{1}={\rm Tr}_{2}\rho_{12}$ for the partial
state in ${\cal H}_{1}$. If $\Phi $ is the channel in ${\cal H}_{1}$, we
denote $\rho_{12}^{\prime}=(\Phi \otimes {\rm Id})(\rho_{12})$. Let us
dilate the channel $\Phi $ to a unitary evolution $U_{13}$ with the
environment ${\cal H}_{3}$, initially in a pure state $\rho
_{3}=|\phi_{3}\rangle \langle \phi_{3}|$. The the final state of the
environment is 
\[
\rho_{3}^{\prime}={\rm Tr}_{1}U_{13}(\rho_{1}\otimes |\phi_{3}\rangle
\langle \phi_{3}|)U_{13}^{\ast }\equiv \Psi (\rho_{1}). 
\]
Since the state of the composite system ${\cal H}_{1}\otimes {\cal H}
_{2}\otimes {\cal H}_{3}$ remains pure after the unitary evolution, its
partial states $\rho_{12}^{\prime},\rho_{3}^{\prime}$ are isometric $\cite
{lin}$. Therefore 
\[
||(\Phi \otimes {\rm Id)(\rho_{12})}||_{p}
=||\rho_{12}^{\prime}||_{p}=||\rho _{3}^{\prime}||_{p}=||\Psi
(\rho_{1})||_{p}. 
\]
Now the map $\rho_{1}\rightarrow \Psi (\rho_{1})$ is affine and the norm is
convex, therefore the maximum of the quantity above is attained on pure $
\rho_{1},$ whence $\rho_{12}=\rho_{1}\otimes \rho_{2},$ and the statement
follows. $\Box $

We shall specifically need this Lemma in the case $p=\infty$. It is
instructive to see an alternative direct proof in this case.

\noindent {\it Proof. } In what follows we take $\rho =|\phi \rangle \langle
\phi |$. We compute the operator norm of the Hermitian operator $\Phi (\rho
) $ as $\Vert \Phi (\rho )\Vert =\sup_{\psi }\langle \psi ,\Phi (\rho )\psi
\rangle $, and take $\Phi $ to be given in the Kraus decomposition (\ref{kra}
). Then 
\[
\puri(\Phi )=\sup_{\phi ,\psi }\sum_{k}\langle \psi ,A_{k}\phi \rangle
\langle \phi ,A_{k}^{\ast }\psi \rangle \;, 
\]
where the supremum is over all unit vectors in the appropriate spaces. The
expression under the supremum can be read as the $\ell ^{2}$-norm of a
vector with components $\langle \phi ,A_{k}^{\ast }\psi \rangle $. We write
this norm also as ``the largest scalar product with a unit vector'' $\chi $,
i.e., 
\begin{eqnarray}
\puri(\Phi ) &=&\left( \sup_{\phi ,\psi ,\chi }\sum_{k}\overline{\chi_{k}}
\langle \phi ,A_{k}^{\ast }\psi \rangle \right) ^{2}=\left( \sup_{\psi ,\chi
}\Vert \sum_{k}\overline{\chi_{k}}A_{k}^{\ast }\psi \Vert \right) ^{2} \\
&=&\sup_{\chi }\Vert \sum_{k}\overline{\chi_{k}}A_{k}^{\ast }\Vert
^{2}=\sup_{\chi }\Vert \sum_{k}\chi_{k}A_{k}\Vert ^{2}\;,
\end{eqnarray}
where all suprema are over unit vectors. Obviously, the Kraus operators for $
\Phi \otimes {\rm Id}$ are $A_{k}\otimes {I}$, so 
\[
\puri(\Phi \otimes {\rm Id})=\sup_{\chi }\Vert \sum_{k}\chi
_{k}(A_{k}\otimes I)\Vert ^{2}=\sup_{\chi }\Vert \left( \sum_{k}\chi
_{k}A_{k}\right) \otimes I\Vert ^{2}=\puri(\Phi )\;. 
\]
$\Box $

We will also need the analogous result for a quantity in which the two
vectors $\phi ,\psi $ in the above proof are fixed to be the same: for any
channel $\Phi $, let 
\begin{equation}
\pur\flat(\Phi )=\sup_{\psi }\bigl\langle\psi ,\Phi (|\psi \rangle \langle
\psi |)\psi \bigr\rangle\;,
\end{equation}
where the supremum is again over all unit vectors. Note that this expression
only makes sense, if the channel does not change the type of system, i.e.,
input and output algebra are the same. Then 
\begin{equation}
\pur\flat(\Phi \otimes {\rm Id})=\pur\flat(\Phi )\;.  \label{purflat+}
\end{equation}

\noindent {\it Proof:\ }\ Again we use Kraus decomposition (\ref{kra}). For $
\psi $ we use the Schmidt decomposition $\psi =\sum_{\mu }\sqrt{c_{\mu }}\
e_{\mu }\otimes e_{\mu }^{\prime }$, where the $e_{\mu }$ and 
$e_{\mu }^{\prime}$ are
orthonormal systems. Then the expression to maximized on the left-hand side
becomes 
\begin{eqnarray*}
&&\sum_{\mu \nu \alpha \beta k}(c_{\mu }c_{\nu }c_{\alpha }c_{\beta })^{1/2} 
\bigl\langle e_{\mu }\otimes e_{\mu }^{\prime },(A_{k}\otimes {I})e_{\nu
}\otimes e_{\nu }^{\prime }\bigr\rangle\bigl\langle e_{\alpha }\otimes
e_{\alpha }^{\prime },(A_{k}^{\ast }\otimes {I})e_{\beta }\otimes e_{\beta
}^{\prime }\bigr\rangle \\
&=&\sum_{\mu \nu \alpha \beta k}(c_{\mu }c_{\nu }c_{\alpha }c_{\beta
})^{1/2} \bigl\langle e_{\mu },A_{k}e_{\nu }\bigr\rangle\bigl\langle 
e_{\alpha },A_{k}^{\ast }e_{\beta }\bigr\rangle\ \delta_{\mu \nu
}\delta_{\alpha \beta } \\
&=&\sum_{\mu \alpha k}c_{\mu }c_{\alpha }\ \bigl\langle e_{\mu },A_{k}e_{\mu
}\bigr\rangle\bigl\langle e_{\alpha },A_{k}^{\ast }e_{\alpha }\bigr\rangle 
=\sum_{k}{\rm tr}(\rho_{1}A_{k}){\rm tr}(\rho_{1}A_{k}^{\ast })\;,
\end{eqnarray*}
where $\rho_{1}=\sum_{\mu }c_{\mu }|e_{\mu }\rangle \langle e_{\mu }|$ is
the reduced density matrix belonging to $\psi $. Since the function $\rho
_{1}\mapsto |{\rm tr}(\rho_{1}A_{k})|^{2}$ is convex, this expression
attains its maximum with respect to $\psi $ when $\rho_{1}$ is pure, i.e.,
when $\psi $ is a product. $\Box $

\section{Weak Noise}

\label{weaknoise} One testing ground for the multiplicativity/additivity
conjecture are channels close to the identity. For such channels the purity
parameters can be evaluated in lowest order in the deviation from the
identity. Doing this for each subchannel and for their tensor product, one
can explicitly check the conjecture. As the following result shows, this
test supports the conjecture.

Consider a channel with weak noise, i.e. choose some channel $\Phi$ on $
{\cal B(H)}$, and set 
\begin{equation}
\Phi ^{(\epsilon )}=(1-\epsilon ){\rm Id}+\epsilon \Phi \;.  \label{lownoise}
\end{equation}
For small $\epsilon $ this is a weak noise channel, which has the property
that for {\it any} pure input the output will be nearly pure.

{\bf Theorem. }The multiplicativity hypothesis for \ the quantities $
\pur p
(\Phi )$ with $1\leq p\leq \infty $ and the additivity \ hypothesis for \
the quantity $\purS(\Phi )$ hold true\ approximately in the leading order in 
$\epsilon $.

{\it Proof.} In order to estimate these quantities for the weak noise
channels, we need to estimate entropy, and the $p$-norms near a pure state.
Let $\rho $ be a density operator on a $d$-dimensional Hilbert space, and
suppose that $\Vert \rho \Vert =1-\epsilon +{\bf o}(\epsilon )$. Then the
leading order of the other norms is determined completely by $\epsilon $: 
\begin{eqnarray}
\Vert \rho \Vert_{p} &=&1-\epsilon +{\bf o}(\epsilon )\;\quad 
\hbox{for
$p>1$} \\
H(\rho ) &=&-\epsilon \log \epsilon +{\bf o}(\epsilon \log \epsilon )\;,
\end{eqnarray}
where as usual ${\bf o}(\epsilon )$ stands for terms going to zero faster
than $\epsilon $ as $\epsilon \rightarrow 0$. In this case we can say more:
in first line we have $0\leq \hbox{remainder}\leq C\;\epsilon ^{p}$, for $
\epsilon <1/2$, where $C$ is a constant depending only on the dimension.
Similarly, the estimates in the second line are independent of the details
of $\rho $. Hence in leading order all the variational expressions are
equivalent: each one amounts to maximizing $\epsilon $.

Let us go back to the weak noise channel (\ref{lownoise}). To get high
fidelity we need to maximize the leading term, so we can take $\eta =\xi $
in the following computation: 
\begin{eqnarray}
\puri(\Phi ^{(\epsilon )}) &=&\sup_{\xi ,\eta }\langle \xi ,\Phi ^{(\epsilon
)}(|\eta \rangle \langle \eta |)\xi \rangle  \nonumber \\
&=&\sup_{\xi ,\eta }\Big((1-\epsilon )|\langle \xi ,\eta \rangle
|^{2}+\epsilon \langle \xi ,\Phi (|\eta \rangle \langle \eta |)\xi \rangle 
\Big)  \nonumber \\
&=&1-\epsilon +\epsilon \sup_{\xi }\langle \xi ,\Phi (|\xi \rangle \langle
\xi |)\xi \rangle \ +{\bf o}(\epsilon )  \nonumber \\
&=&1-\epsilon +\epsilon \pur\flat(\Phi )+{\bf o}(\epsilon )\;.
\label{puri@lownoise}
\end{eqnarray}
Note that in all these estimates the remainder estimates can be done
uniformly for all channels, depending only on dimension.

A tensor product of weak noise channels (\ref{lownoise})\ is again of the
same form: 
\begin{eqnarray*}
\Phi ^{(\epsilon )} &=&\Phi_{1}^{(\epsilon )}\otimes \cdots \Phi
_{n}^{(\epsilon )} \\
&=&(1-\epsilon )^{n}{\rm Id}+\epsilon (1-\epsilon )^{n-1}\left( \Phi
_{1}\otimes {\rm Id}_{2\dots n}+\cdots +{\rm Id}_{1\dots n-1}\otimes \Phi
_{n}\right) +{\bf o}(\epsilon ) \\
&=&(1-n\epsilon ){\rm Id}+n\epsilon \;\delta \Phi +{\bf o}(\epsilon )\;,
\end{eqnarray*}
where $\delta \Phi $ is the average of the $n$ channels ${\rm Id}_{1\dots
k-1}\otimes \Phi_{k}\otimes {\rm Id}_{k+1\dots n}$. Hence, in order to
compute the leading order of $\puri(\Phi ^{(\epsilon )})$ by formula (\ref
{puri@lownoise}) we have to determine $\pur\flat(\delta \Phi )$. We have 
\begin{eqnarray*}
\frac{1}{n}\sum_{k=1}^{m}\pur\flat(\Phi_{k}) &\leq &\pur\flat(\delta \Phi )
\\
&\leq &\frac{1}{n}\sum_{k=1}^{m}\pur\flat(\Phi_{k}\otimes {\rm Id}) \\
&=&\frac{1}{n}\sum_{k=1}^{m}\pur\flat(\Phi_{k})\;
\end{eqnarray*}
where we have used in turn: insertion of product states into the supremum
defining $\pur\flat(\delta \Phi )$, convexity of $\pur\flat()$ as a supremum
of affine functionals, and finally the restricted additivity result (\ref
{purflat+}). Hence equality holds, which means that in the leading order in $
\epsilon $ all the variational expressions for the purity quantities $\pur*(
) $, with $\ast =p,H,\flat $ are attained at product states. $\Box $

\section{Depolarizing Channels}

\label{sec:dep}

A channel is called {\it bistochastic} if $\Phi (I)=I$, where $I$ is the
unit operator in ${\cal B(H)}$. An important example is the {\it depolarizing
} channel \cite{ben} 
\[
\Phi (\rho )=(1-p)\rho +\frac{p}{d}(\mbox{Tr}\rho )I,\ \rho \in {\cal B(H)}
,\ 0<p<1, 
\]
where $d=\mbox{dim}{\cal H}$. A channel is called {\it binary} if $d=2$.

{\bf Lemma.} Let $\Phi $ be binary bistochastic channel, then 
\begin{equation}
C(\Phi )=\log 2-\min_{\rho \in {\cal S(H)}}H(\Phi (\rho )).  \label{bbc}
\end{equation}
If $\Phi_{i}$ are binary bistochastic channels, then (\ref{ent}) implies (
\ref{add}).

{\it Proof. }The $\leq $ part of (\ref{bbc}) is obvious from the fact that
for any channel 
\begin{equation}
C(\Phi )\leq \log \mbox{dim}{\cal H}-\min_{\rho \in {\cal S(H)}}H(\Phi (\rho
)),  \label{leq}
\end{equation}
so we need to prove only $\geq $ part. Since the entropy is convex, the
minimum is achieved at the set of extreme points of ${\cal S(H)}$ which is $
{\cal P}_{1}{\cal (H)}$. Let $\rho $ be the minimum point, then taking
equiprobably $\rho_{0}=\rho ,\rho_{1}=I-\rho $, we obtain 
\[
C(\Phi )\geq H(\frac{1}{2}\Phi (I))-\frac{1}{2}[H(\Phi (\rho ))+H(\Phi
(I-\rho ))]. 
\]
Since the channel is bistochastic, this is equal to $H(\frac{1}{2}I)-\frac{1 
}{2}[H(\Phi (\rho ))+H(I-\Phi (\rho ))]$, and since it is binary, this is
equal to the right-hand side of (\ref{bbc}).

To prove the second statement, it is sufficient to prove the $\leq$ part of (
\ref{add}), since $\geq$ part follows from the definitions. But this follows
from (\ref{leq}) and (\ref{bbc}). $\Box$

In the paper \cite{brus} the relation(\ref{add}) was proven for the two
binary depolarizing channels $\Phi_{1},\Phi_{2}$. The proof heavily uses
Schmidt decomposition and as such does not generalizes to the case $n>2$.
The main difficulty is evaluating the entropy of the product channel.
However, it appears to be possible to check the additivity in the limiting
cases of ``weak'' and ``strong'' depolarization in the leading order. Let us
consider a collection of depolarizing channels $\{\Phi_{i}\}$ in the Hilbert
spaces ${\cal H}_{i}$, with parameters $p_{i},d_{i},\ i=1,2,\ldots,n$, and
denote $\Phi =\otimes_{i=1}^{n}\Phi_{i}$, ${\cal H}=\otimes_{i=1}^{n}{\cal H}
_{i}$, $d=\prod_{i=1}^{n}d_{i}$. In the following we shall use symbols $I_{i}
$ and $I=\otimes_{i=1}^{n}I_{i}$ for the identity operators in ${\cal H}_{i}$
and ${\cal H}$ respectively. Let $\epsilon_{L}$ be the tensor product $\phi
_{1}\otimes \ldots\otimes \phi_{n}$, where $\phi_{i}(\rho )=\frac{1}{d_{i}} 
\mbox{Tr}(\rho )I_{i},\ i\in L\subset \{1,2,\ldots,n\}$, and $\phi_{i}(\rho
)=\rho $ otherwise, $\rho \in {\cal S(H}_{i}{\cal )}$. Then $\epsilon_{L}$
is  a conditional expectation onto the subalgebra ${\cal M}_{L}$,
generated by operators of the form $A_{1}\otimes \ldots\otimes A_{n}$, where 
$A_{i}=I_{i}$ for $i\in L$, and is normalized partial trace with respect to its
commutant. It has the property $\min\limits_{P\in {\cal P}({\cal M}_{L})} 
{\rm dim}P=\prod_{i=1}^{n}d_{i}^{\theta_{L}(i)}$, where $\theta_{L}(i)=1$ if 
$i\in L$ and $\theta_{L}(i)=0$ otherwise. Here we denoted by ${\cal P}( 
{\cal M}_{L})$ the set of all orthogonal projections in ${\cal M}_{L}$.
Notice that the inclusion ${\cal M}_{L_{1}}\subset {\cal M}_{L_{2}}$ holds
if $L_{2}\subset L_{1}$. So $\epsilon_{L_{1}}\epsilon_{L_{2}}=\epsilon
_{L_{1}\vee L_{2}}$.

We shall use the expansion 
\begin{equation}
\Phi =\sum\limits_{L}\prod\limits_{i=1}^{n}p_{i}^{\theta
_{L}(i)}(1-p_{i})^{1-\theta_{L}(i)}\epsilon_{L}.  \label{tilde}
\end{equation}
Weak (strong) depolarization corresponds to the case where all $p_{i}$
(respectively, $1-p_{i}$) are small parameters, which we assume to be of the
same order. 

{\bf Proposition}{\sc . }The relation (\ref{add}) holds in the cases of weak
and strong depolarization approximately in the leading order.

{\it Proof. }In the case of weak depolarization the statement follows from
the Theorem in Section~\ref{weaknoise}.

In the case of strong depolarization we have to retain all the terms up to
the second order in $q_{i}=1-p_{i}$. Then the leading terms are 
\[
\Phi (P)\sim d^{-1}\left[ I+\sum\limits_{i=1}^{n}q_{i}(d_{i}P_{i}-I)+\sum_{1
\leq i<j\leq n}q_{i}q_{j}(1-d_{i}P_{i}-d_{j}P_{j}+d_{i}d_{j}P_{ij})\right] , 
\]
where we denoted $d=\prod_{i=1}^{n}d_{i}$, $P_{i}$ is the partial state of $
P $ in the $i-$th Hilbert space, multiplied by the unit operator in the
tensor product of the remaining Hilbert spaces, and similarly $P_{ij}$.
Denoting the first (second) sum in the squared brackets $A_{1}$
(respectively $A_{2}$ ) one easily sees that both are traceless operators.
Moreover, up to the second order, 
\[
\Phi (P)\log \Phi (P)\sim d^{-1}\left[ (1+A_{1}+A_{2})\log d^{-1}+\left(
A_{1}+A_{2}+\frac{A_{1}^{2}}{2}\right) \right] 
\]
and 
\[
H(\Phi (P))\sim \log d-\frac{{\rm Tr}A_{1}^{2}}{2d}. 
\]
But 
\[
{\rm Tr}A_{1}^{2}=d\sum\limits_{i=1}^{n}q_{i}^{2}(d_{i}{\rm Tr}\rho
_{i}^{2}-1), 
\]
where $\rho_{i}$ is the partial state of $P$ in the $i-$th Hilbert space,
which is maximized if and only if $\rho_{i}$ is one-dimensional projection,
i.e. $P=\rho_{1}\otimes \dots \otimes \rho_{n}.\Box $

Partial answers to the multiplicativity hypothesis are given by the
following Theorem. In fact, multiplicativity of $\puri(\Phi )$ for binary
bistochastic maps follows from a more general result in \cite{rusk}.

{\bf Theorem.} 
\[
\begin{array}{llll}
(i)\quad & \pur{2}(\Phi ) & = \prod\limits_{i=1}^{n}\pur{2}({\Phi } _{i}) & 
= \prod\limits_{i=1}^{n}\left( \frac{d_{i}-1}{d_{i}}(1-p_{i})^{2}+\frac{1 }{
d_{i}}\right)^{1/2} , \\ 
(ii)\quad & \puri(\Phi ) & = \prod\limits_{i=1}^{n}\puri(\Phi _{i}) & =
\prod\limits_{i=1}^{n}\left( 1-\frac{p_{i}(d_{i}-1)}{d_{i}}\right) , \\ 
(iii)\quad & \pur{-\infty}(\Phi ) & = \prod\limits_{i=1}^{n}\pur{-\infty}({
\Phi } _{i}) & = \prod\limits_{i=1}^{n}\frac{p_{i}}{d_{i}}.
\end{array}
\]

{\it Proof. }$(i)$ It follows from the relation $\min\limits_{P\in {\cal P}( 
{\cal M}_{L})}\mbox{dimP}=\prod_{i=1}^{n}d_{i}^{\theta_{L}(i)}$ that 
\[
\mbox{Tr}(\epsilon_{L_{1}}(P)\epsilon_{L_{2}}(P))=\mbox{Tr}(P\epsilon
_{L_{1}\vee L_{2}}(P))=\mbox{Tr}(\epsilon_{L_{1}\vee L_{2}}(P)^{2})\leq
\prod_{i=1}^{n}d_{i}^{-\max (\theta_{L_{1}}(i),\theta_{{L_{2}}}(i))} 
\]
for arbitrary $P\in {\cal P}_{1}{\cal (H)}$ and equality holds only for
factorizable projections. Hence 
\[
\mbox{Tr}((\Phi (P))^{2})=\mbox{Tr}((\sum\limits_{L}\prod
\limits_{i=1}^{n}p_{i}^{\theta_{L}(i)}(1-p_{i})^{1-\theta_{L}(i)}\epsilon
_{L}(P))^{2}) 
\]
\[
\leq \sum\limits_{L_{1},L_{2}}\prod\limits_{i=1}^{n}p_{i}^{\theta
_{L_{1}}(i)+\theta_{L_{2}}(i)}(1-p_{i})^{2-\theta_{L_{1}}(i)-\theta
_{L_{2}}(i)}d_{i}^{-\max (\theta_{L_{1}}(i),\theta_{{L_{2}}}(i))}= 
\]
\[
=\prod\limits_{i=1}^{n}\sum_{\theta_{1},\theta_{2}=0,1}p_{i}^{\theta
_{1}+\theta_{2}}(1-p_{i})^{2-\theta_{1}-\theta_{2}}d_{i}^{-\max (\theta
_{1},\theta_{2})}=\prod\limits_{i=1}^{n}\left( \frac{d_{i}-1}{d_{i}}
(1-p_{i})^{2}+\frac{1}{d_{i}}\right) . 
\]

$(ii)$ Let us estimate 
\[
||\Phi (P)||\leq \sum\limits_{L}\prod\limits_{i=1}^{n}p_{i}^{\theta
_{L}(i)}(1-p_{i})^{1-\theta_{L}(i)}||\epsilon_{L}(P)|| 
\]
\[
\leq \sum\limits_{L}\prod\limits_{i=1}^{n}\left( \frac{p_{i}}{d_{i}}\right)
^{\theta_{L}(i)}(1-p_{i})^{1-\theta_{L}(i)}=\prod\limits_{i=1}^{n}\left( 1- 
\frac{p_{i}(d_{i}-1)}{d_{i}}\right) . 
\]
which proves the second statement.

$(iii)$ We have 
\begin{equation}
||\Phi (P))^{-1}||\leq \sum\limits_{k=0}^{+\infty }||\Phi (I-P)||^{k}.
\label{*}
\end{equation}
Let us calculate 
\[
\max\limits_{P\in {\cal P}_{1}({\cal B(H)})}||\Phi (I-P)||\leq
\max\limits_{Q\in {\cal P}_{d-1}({\cal B(H)})}||\Phi (Q)||. 
\]
We have 
\[
||\Phi (Q)||\leq
\prod\limits_{i=1}^{n}p_{i}||\epsilon_{\{1,2,\ldots,n\}}(Q)|| 
\]
\[
+\sum\limits_{L\neq \{1,2,\ldots,n\}}\prod\limits_{i=1}^{n}p_{i}^{\theta
_{L}(i)}(1-p_{i})^{1-\theta_{L}(i)}||\epsilon_{L}(Q)||\leq 
\]
\begin{equation}
\left( 1-d^{-1}\right) \prod\limits_{i=1}^{n}p_{i}+\sum\limits_{L\neq
\{1,2,\ldots,n\}}\prod\limits_{i=1}^{n}p_{i}^{\theta
_{L}(i)}(1-p_{i})^{1-\theta_{L}(i)}=1-\prod\limits_{i=1}^{n}\frac{p_{i}}{
d_{i}}.  \label{**}
\end{equation}
Here we have used the equality $\sum\limits_{L}\prod\limits_{i=1}^{n}p_{i}^
{\theta_{L}(i)}(1-p_{i})^{1-\theta_{L}(i)}=1$. Substituting (\ref{**}) into (
\ref{*}), we get the last statement. $\Box $

{\bf Acknowledgment}. The second author (ASH) acknowledges the support of the
A. von Humboldt Foundation.


\begin{thebibliography}{9}
\bibitem{smo}  C. H. Bennett, C. Fuchs, J. A. Smolin, Entanglement enhanced
classical communication on a noisy quantum channel, in: Proc. 3d Int. Conf.
on Quantum Communication and Measurement, ed. by C. M. Caves, O. Hirota, A.
S. Holevo, Plenum, NY 1997. LANL e-print quant-ph/9611006.

\bibitem{ben}  C. H. Bennett, P. W. Shor, Quantum information theory, IEEE
Trans. on Inform. Theory, {\bf IT-44}, 2724-2742, 1998.

\bibitem{brus}  D. Bruss, L. Faoro, C. Macchiavello, M. Palma, Quantum
entanglement and classical communication through a depolarizing channel.
J. Mod. Opt. {\bf 47}, 325-332, 2000. LANL e-print quant-ph/9903033.

\bibitem{fuchs}  C. Fuchs, private communication.

\bibitem{hol}  A.S. Holevo, Quantum coding theorems, Russian Math. Surveys
{\bf 53:6}, 1295-1331, 1998. LANL e-print quant-ph/9808023.

\bibitem{rusk}  C. King, M. B. Ruskai, Minimal entropy of states emerging
from noisy quantum channels. LANL e-print quant-ph/9911079.

\bibitem{lin}  G. Lindblad, Quantum entropy and quantum measurements, in:
Proc. Int. Conf. on Quantum Communication and Measurement, ed. by C.
Benjaballah, O. Hirota, S. Reynaud, Lect. Notes Phys. {\bf 378}, 71-80,
Springer-Verlag, Berlin 1991.

\bibitem{sch}  B. Schumacher, M. D. Westmoreland, Optimal signal ensembles.
LANL e-print quant-ph/9912122.
\end{thebibliography}
\end{document}